\documentclass[aps,prb,twocolumn]{revtex4-1}
\usepackage{graphicx}
\usepackage[bottom]{footmisc}
\usepackage{amsmath}
\usepackage{amssymb}
\usepackage{braket}
\usepackage{hyperref}
\usepackage{subfigure}
\usepackage[usenames,dvipsnames]{color}
\usepackage{graphicx}
\usepackage{subfigure}
\usepackage{cleveref}
\usepackage{amsmath,amssymb}

\newcommand{\ct}{\cite}

\newcommand{\bi}{\bibitem}
\newcommand{\be}{\begin{equation}}
\newcommand{\ee}{\end{equation}}
\newcommand{\ba}{\begin{eqnarray}}
\newcommand{\ea}{\end{eqnarray}}

\begin{document}
\begin{@twocolumnfalse}

\title{Dynamical preparation of a topological state and out of equilibrium bulk boundary correspondence in a SSH chain under periodic driving}
\author{Souvik Bandyopadhyay} 
\email{souvik@iitk.ac.in}
\author{Amit Dutta}
\affiliation{Department of Physics, Indian Institute of Technology, Kanpur, Kanpur 208016, India}

	\begin{abstract}
	
	Exploiting  the possibility of temporal variation of the winding number, we have prepared a SSH chain in its {\it stroboscopic} topological state, starting from the trivial one, by application of
	a periodic perturbation. 
	The periodic driving, we employ here, is adiabatically switched on to break the particle-hole symmetry and generate a chiral mass term in the effective Floquet Hamiltonian; consequently   the Floquet Hamiltonian also gets deformed without crossing the gapless quantum critical point. The particle hole symmetry is subsequently restored in the Floquet Hamiltonian by adiabatically switching off a part of the periodic potential. Thereafter, the  Floquet Hamiltonian develops a symmetry protected non-trivial topological winding number. 
	Furthermore, we also observe stroboscopic topologically protected localised edge states  in a long open chain and show that a bulk boundary correspondence survives a unitary non-equilibrium situation in 1D BDI Hamiltonians. Moreover, considering an extended SSH chain with higher neighbour hoppings, we dynamically prepare the system in a stroboscopic out-of-equilibrium topological insulator state starting from a metallic regime. At the same time, we establish the dynamical preparation of higher winding phases in an extended SSH chain with stroboscopic bulk-boundary correspondence in the non-equilibrium state of the system.

	\end{abstract}
	\maketitle
\end{@twocolumnfalse}

\section{Introduction}
Preparing a system in its topologically protected state is of great interest both theoretically\cite{kitaev01,kane05,bernevig06,fu08,zhang08,sato09,sau10a,sau10b,lutchyn10,oreg10} and experimentally \ct{mourik12,rokhinson12,deng12,das12,churchill13,finck13,alicea12,leijnse12,beenakker13,stanescu13} as well as from the viewpoint of quantum information theoretical studies \ct{kitaev16,pachos17} (for review see, [\onlinecite{moore10,shen12,bernevig13}]). Topological
and Chern insulators are novel phases of matter which go beyond the conventional Landau Ginzberg paradigm and are characterised by a bulk topological invariant. The topological properties
of these systems are protected by a gap in the bulk spectrum and are  manifested in  the occurrence of  robust edge states. The so called {\it topological} invariant, characterising the topological phase,
can undergo a change only when the system is tuned across a gapless (topological) quantum critical point (QCP) \ct{shen12,bernevig13}.

 One of the possible ways of realising the topological phases is through the application of a time periodic global driving of a topologically trivial initial state; these periodic perturbations lead to the realisation of new topological phases of matter which have no equilibrium counterparts \cite{oka09,kitagawa11,rudner13,balseiro14,mitra15,gil16} for example,
Floquet graphene \ct{oka09,kitagawa11}, topological insulators \ct{lindner11,cayssol13} and  edge Majoranas \ct{thakurathi13,kundu13}. 
In the above studies, the effective Hamiltonian dictating the non-equilibrium dynamics of the system become topologically non-trivial despite the initial static Hamiltonian being trivial; although the evolved
state is not necessarily topological. The dynamical preparation of a system in an out-of-equilibrium Floquet state is a topic of ongoing and extensive field of study. Therefore, the success of such novel applications depends not only on the dynamical generation of a topological Hamiltonian but also a topologically non-trivial dynamical state.
Interestingly, for two-dimensional (2D) Chern insulating systems obeying periodic boundary  conditions,  it has been shown that it is not possible to prepare a non-trivial topological state via unitary evolution starting from a trivial initial state; this is a consequence of
the invariance of   the bulk Chern number.  However, for a system with boundaries, a new topological invariant i.e. the Bott index has been shown to temporally vary under unitary dynamics \cite{rigol15}. Also, the edge states of such 2D systems can exhibit non-trivial behavior despite Chern triviality of the non-equilibrium bulk state \cite{cooper15,utso17,sougato18}. Hence, the familiar notion of the bulk-boundary correspondence breaks down.

Recently,  it has been established that unlike the Chern number  the dynamical counter part of the winding number may change under unitary dynamics \ct{foster13,foster14,ginley18,bandyopadhyay19}. However, many questions remain unexplored. Referring to studies  presented in Ref. \onlinecite{bandyopadhyay19}, the first question that arises is whether the ``dynamical'' winding number can be tuned to an integer quantised  value so that the system dynamically  reaches a state characterised by an appropriate topological index.  While the previous work  \ct{bandyopadhyay19} deals with closed systems with a periodic boundary condition (PBC), a far more interesting situation may arise  when a similar dynamics is studied with open boundary conditions (OBC). The relevant question would be whether the change in the winding number is manifested in the occurrence of  zero-energy edge states in the corresponding
OBC situation, i.e., whether there exists a ``non-equilibrium bulk boundary correspondence".  When {\it both} the above conditions, namely the integer quantised winding number and the occurrence of zero-energy
edge states are achieved,  there emerges a possibility of  dynamical engineering of  1D topological phases  through unitary driving protocols.

In this work, we  {precisely address these two questions:} We explore the possibility of engineering a topological state and the fate of the related bulk-boundary correspondence in a non-equilibrium 1D SSH chain under a time-periodic unitary drive.
	Starting from the topologically trivial ground state of a SSH chain \ct{shen12}, which is the simplest topological insulator belonging to the BDI symmetry class \ct{altland97}, we subject the state to a time periodic drive which is adiabatically switched on. For sufficiently high frequencies, the Floquet Hamiltonian governing the dynamics at stroboscopic instances, develops a chiral mass which explicitly breaks the particle-hole symmetry  and opens up a gap in the bulk spectrum\ct{kitagawa11}. In the presence of the staggered mass, the parameters of the  Floquet Hamiltonian also makes a transition from an initial topologically trivial value to those corresponding to the topologically
non-trivial phase. The periodic potential generating the chiral mass is adiabatically switched off subsequently to restore all the protecting symmetries in the Floquet Hamiltonian. The adiabatic deformation of the Floquet Hamiltonian allows the temporal state of the system to dynamically follow the instantaneous ground state of the Floquet Hamiltonian (in the high frequency limit) developing a non-trivial winding number. Hence, we propose a novel way to prepare topologically non-trivial states hosting robust edge modes in a one-dimensional model.  {To the best of our knowledge, our work is the first
one that not only generates an effective topological Hamiltonian but also prepares the system in a non-equilibrium topological state  and at the same time establishing an out-of-equilibrium  bulk-boundary correspondence.\\}

\quad {To further establish the efficacy of our unitary preparation protocol, we  have studied the unitary preparation of a topological state in an extended version of the SSH model with higher neighbor hopping. In equilibrium, this model allows for topological phases with higher winding number and a non-topological metallic phase depending on the parameter values. What we show are the following:
(i) The strength of the next nearest neighbor hopping can be made to vanish (consequently restoring all BDI symmetries) under a completely periodic drive  {with a linearly ramped amplitude.} This eventually prepares the out-of-equilibrium system in a stroboscopic topological insulator state with associated edge modes starting from a  metallic state (with an indirect band-gap) of the AI symmetric extended chain. (ii) Starting from a situation, in which the next nearest neighbor hopping term is zero in the initial Hamiltonian, we show that using the similar periodic protocol (again with a linearly ramped amplitude), we can prepare the system in a topological stroboscopic state with winding number changed to two from the initial equilibrium value of unity. Correspondingly, we also establish the dynamical emergence of two pairs of stroboscopic edge modes in the out-of-equilibrium state of the system. These zero quasi-energy modes are localized at different sites near the boundary of a driven chain under open boundary conditions. This study further establishes the importance of studying the out-of-equilibrium unitary preparation of a Floquet topological state and an associated bulk boundary correspondence.\\}

\quad {The paper is organised in the following sections: In Sec.~\ref{sec:SSH_intro}, we provide a pedagogical introduction to the SSH chain and its associated symmetries followed by its equilibrium topological characterisation. We also briefly discuss the fate of the topological properties under non-equilibrium unitary dynamics in such systems. In Sec.~\ref{sec: protocol} we explore the possibility of out-of-equilibrium quantisation of the dynamical winding number with respect to a unitary driving protocol. We propose a periodic driving protocol to prepare a stroboscopic topological state with a non-trivial stroboscopic winding number. In Sec.~\ref{sec: OBC}, we  study the non-equilibrium stroboscopic behaviour of the SSH chain with OBCs under the   {same} periodic driving protocol introduced in Sec.~\ref{sec: protocol}. We numerically establish the existence of localized edge states in the stroboscopic state of the system. To further probe the dynamical emergence of bulk-boundary correspondence, we study the topological properties of a non-equilibrium extended SSH Hamiltonian in Sec.~\ref{sec: extendedSSH}. We begin the section with a brief  introduction to  the SSH chain with extended second and third nearest neighbour hoppings, associated symmetries and its topological phases. In subsection \ref{sec: SecondHopping}, we introduce a unitary driving protocol to stroboscopically restore all BDI symmetries in the otherwise AI symmetric extended SSH chain. This restores the topological bulk boundary correspondence in the stroboscopic state of the system and the bulk of the system makes a transition from a metallic phase to a stroboscopic topological insulator phase. In subsection \ref{winding2}, starting from a BDI symmetric topological phase of the extended SSH model, we adiabatically ramp a time-periodic perturbation to prepare the system in a topological state having a higher value of the topological invariant. We also establish the associated emergence of a stroboscopic bulk boundary correspondence through the existence of two pairs of stroboscopic localised edge modes in the out-of equilibrium state. Lastly, in Sec.~\ref{conclusion}, we conclude with a brief summary of our results and draw conncetions with existing experimental set-ups and previous experimental studies. }


\section{The Su-Schrieffer-Heeger (SSH) Model and its symmetries~}
\label{sec:SSH_intro}
The Hamiltonian for the SSH model,  that describes a $1$D lattice having a two atom sublattice structure with $\mathcal{N}$ unit cells, can be written in terms of the (spin polarised) fermionic creation and annihilation operators as,
\begin{equation}\label{eq:H}
H=\sum_{n=1}^{\mathcal{N}}(vc^{\dagger}_{n,1}c_{n,2}+wc^{\dagger}_{n,2}c_{n+1,1}+h.c).
\end{equation}
Here, $h.c.$ denotes the hermitian conjugate,
$v$ and $w$ are the intra-unit-cell  and inter-unit-cell hopping amplitudes, respectively, and subscript
 ${n,i}$  denotes the sublattice position $i~(i=1,2)$ of the $n^{th}$ unit cell. 
After performing a tight-binding analysis one can write the Hamiltonian in Eq.~\eqref{eq:H} as,
\begin{equation}\label{eq:Hk}
H_0(k)=\bigoplus_{k}\vec{h}(k).\vec{\sigma},
\end{equation} 
where, 
\begin{eqnarray}\label{eq:hk}
\nonumber h_{x}(k)&=& Re(v)+|w|\cos(k+arg(w))\\
\nonumber h_{y}(k)&=& -Im(v)+|w|\sin(k+arg(w))\\
\noindent h_{z}(k)&=& 0.
\end{eqnarray}
The ground state  $ |\psi_{k
 }\rangle$ of the Hamiltonian $H_0(k)$ is characterised by the following topological winding number:
\begin{equation}\label{eq:winding}
\nu=\frac{i}{2\pi}\int^{\pi}_{-\pi}dk \frac{d}{dk}ln(h_{x}+ih_{y})=\frac{i}{2\pi }\int_{BZ}\langle\psi_{k}|\partial_{k}|\psi_{k}\rangle dk,
\end{equation}
which is quantized and can only assume integral values. 
If $|v| > |w|$, $\nu =0$, the chain is in the trivial phase; on the other hand, 
$\nu=1$  if $|v| <|w|$ when the chain hosts topologically protected robust end states. 
The energy gap between the two bands vanishes at  the QCP  $|v| = |w|$, where the winding number becomes undefined. Thus the topologically protected winding number can only change when the parameters are tuned across a gapless point.

We shall focus on a SSH chain belonging to BDI class of non-interacting topological Hamiltonians respecting the time reversal symmetry ($\mathcal{T}$), the particle-hole symmetry ($\mathcal{P}$) and the sublattice symmetry ($\mathcal{S}$). 


{\it Winding number  in a driven system:~}
We consider the temporal evolution of the equilibrium topological invariant, i.e., the winding number under a unitary  drive  with an initial (ground) state
$|\psi_{k}(0)\rangle$,  which evolves under the time-dependent  instantaneous Hamiltonian $H_k(t)$,
\begin{equation}\label{eq:evolve}
\begin{split}
|\psi_{k}(t)\rangle=\mathbb{T} e^{-i\int_{0}^{t}H_{k} (t^{'})dt^{'}}|\psi_{k}(0)\rangle\\ 
\equiv e^{-iH_{k}^{eff}(t) t}|\psi_{k}(0)\rangle\\
=U_{k}(t)|\psi_{k}(0)\rangle.
\end{split}
\end{equation}
Here, $H^{eff}_k(t)$ is a time-dependent effective Hamiltonian acting as a generator of unitary evolution on the driven system and $\mathbb{T}$ is the time ordering operator.
As in ref. [\onlinecite{bandyopadhyay19}], it has been established that the winding number is temporally invariant if the effective Hamiltonian capturing the dynamics of the chain respect the discrete symmetry combinations  
of $\mathcal{T}$ and $\mathcal{P}$ or just $\mathcal{P}$.\\

Interestingly, in the case of a time periodic drive with a period $T$, the stroboscopic variation (measured after a complete period of the drive) of the winding number is proportional to the stroboscopic change in the bulk polarization density of the chain in the interval $\left[(m-1)T,mT\right]$,
\begin{equation}\label{avgJ}
\Delta\nu_{m}=\frac{\nu(mT)-\nu((m-1)T)}{T}=\frac{1}{T}\int_{(m-1)T}^{mT}j(t)dt
\end{equation}

The stroboscopically observed winding number has been established  to depend only on the symmetries of the Floquet Hamiltonian, i.e.,
 \begin{equation}\label{delta nu}
 \nu(mT)=\nu(0)+\frac{i}{2\pi}\oint\bra{\psi_k(mT)}(\partial_{k}e^{-iH_F(k)mT})\ket{\psi_k(0)}dk
 \end{equation}
 where $H_F(k)$ is the Floquet Hamiltonian. If $H_F(k)$ respects either $\mathcal{P}$ or $\mathcal{P}$ and $\mathcal{T}$, the stroboscopic winding number remains dynamically invariant.\\

 However, if the $\mathcal{P}$ symmetry is broken in the Floquet Hamiltonian, the stroboscopic winding number evolves temporally resulting the generation of a bulk polarisation current \ct{bandyopadhyay19} . The variation of the winding number opens up the possibility of the preparation of a Floquet non-trivial topological phase starting from a trivial one.\\


 \section{Dynamical preparation of a non-trivial topological state: PBC}
 \label{sec: protocol}

 Starting from the trivial ground state of a  BDI symmetric  SSH Hamiltonian, we periodically drive
the system. Choosing a sufficiently high driving frequency $(\omega=2\pi/T)$, the periodic perturbation is switched on adiabatically, allowing the instantaneous state of the system to follow the instantaneous ground state of the Floquet Hamiltonian.
 We choose the protocol,
 \begin{equation}
 \begin{split}
H_k(t)= H_0 + \lambda_k (t)=H_0(k)+A_1(t)V_0 \sigma_x + A_2(t) V_k(t),
 \end{split}
 \end{equation}
 with $V_k(t+T)=V_k(t)$ being periodic in time having a time period of $T$. The amplitudes  $A_1(t)$ and $A_2(t)$ are adiabatic ramps which slowly switches on and off the periodic perturbation,
 \begin{equation}\label{protocol}
 \begin{split}
 A_1(t)=A_2(t)= -t/\tau\text{~~for~~} 0\le t \le \tau, \\
 A_1(t) =1 \text{~~and~~} A_2(t) = -\left(1-\frac{t-\tau}{\tau}\right) \text{~~for~~} 	\tau\le t \le 2\tau,\\
 A_1(t) =1 \text{~~and~~} A_2(t) = 0 \text{~~for~~} 	t\ge 2\tau,
 \end{split}
 \end{equation}
  allowing the Floquet Hamiltonian to get adiabatically modified. The periodic perturbation $V_k(t)$ is chosen to be sinusoidal,
  \begin{equation}\label{Vkt}
 V_k(t)=V_1\sin(\omega t)\sigma_x+V_1\cos(\omega t)\sigma_y.
  \end{equation}

We note that in the interval $0 \le t \le \tau$, the periodic perturbation $V_k(t)$ is switched on adiabatically and in the subsequent interval $\tau \le t \le 2\tau$ a part of the periodic drive is switched off, again adiabatically, to restore all the symmetries.\\
Evidently, due to the presence of the linear ramping, the  protocol is not perfectly periodic in the time interval $0 \le t \le 2\tau$. However, as the ramping is much slower than the time period of the periodic perturbation, the complete evolution may be partitioned into the `fast' and the `slow' time variables  \ct{kitagawa11}: the former is the time scale of the high frequency periodic drive and the later represents the time scale of the linearly ramped amplitude. Therefore, for a fixed value of the amplitude, the drive is completely periodic and a Floquet picture holds. \\ 
If the frequency of the periodic drive is much higher than the characteristic bandwidth of the system and the amplitude of the drive is small enough $\left(\mathcal{A}^2/\omega\ll 1, \mathcal{A}~\text{being the {\it effective} amplitude of the drive}\right)$, the drive is off resonant. The effective Floquet Hamiltonian may be approximated just to incorporate the single-photon virtual processes or nearest neighbour hopping in the Floquet Bloch lattice i.e.,
\begin{equation}
H_F(k)\simeq \frac{1}{T}\int_0^T H_k(t) dt+\frac{1}{\omega}\left[H_{-1}(k),H_{+1}(k)\right]+\mathcal{O}\left( \frac{\mathcal{A}^4}{\omega^2}\right),
\label{eq_floquet_ham}
\end{equation}
where,
$H_{\pm 1}(k)=\frac{1}{T}\int_0^T H(t)e^{\pm i\omega t}dt.$

We now further assume that $A(t)$ changes very slowly $(\tau \gg T)$ such that its change over a time period $T$ may be neglected at very high frequencies. Using Eq.~\eqref{eq_floquet_ham},  the effective Floquet Hamiltonian takes the following form,
\begin{equation}
\label{eq:floquet}
H_F^t(k)\simeq H_0(k)-A_1(t)V_0\sigma_x+\frac{\left[A_2(t)V_1\right]^2}{2\omega}\sigma_z ,
\end{equation}

 The slowly varying amplitude can now be interpreted as an adiabatic deformation of the Floquet Hamiltonian $H_F^t(k)$. The magnitude of $V_0$ is so chosen that the term proportional to $\sigma_x$ in the effective Floquet Hamiltonian Eq.~\eqref{eq:floquet} drives the Floquet Hamiltonian across the QCP  to render the final Floquet Hamiltonian topologically non-trivial at time $t=2\tau$.
 
 The diagonal term on the other hand serves two purposes.  It generates a staggerred mass term which opens up a gap between the Floquet eigenstates and keeps the floquet spectrum gapped at all times. This gap in the Floquet spectrum protects the instantaneous ground states against the generation of excitations particularly near the QCP where the gap otherwise vanishes in the absence of the staggered mass. The  mass term also explicitly breaks the $\mathcal{P}$ and $\mathcal{S}$ symmetry at the same time  in the Floquet Hamiltonian $H_F(k)$ causing the bulk topological invariant to change stroboscopically. At the end of the switching off (i.e., at $t=2\tau$), the particle hole symmetry is completely restored in the final Floquet Hamiltonian $H_F^t(k)$ which now belongs to a completely symmetric BDI SSH model in the topologically non-trivial sector. 
 
 \begin{figure}

	\begin{center}
		\includegraphics[width=0.48\textwidth,height=0.65\columnwidth]{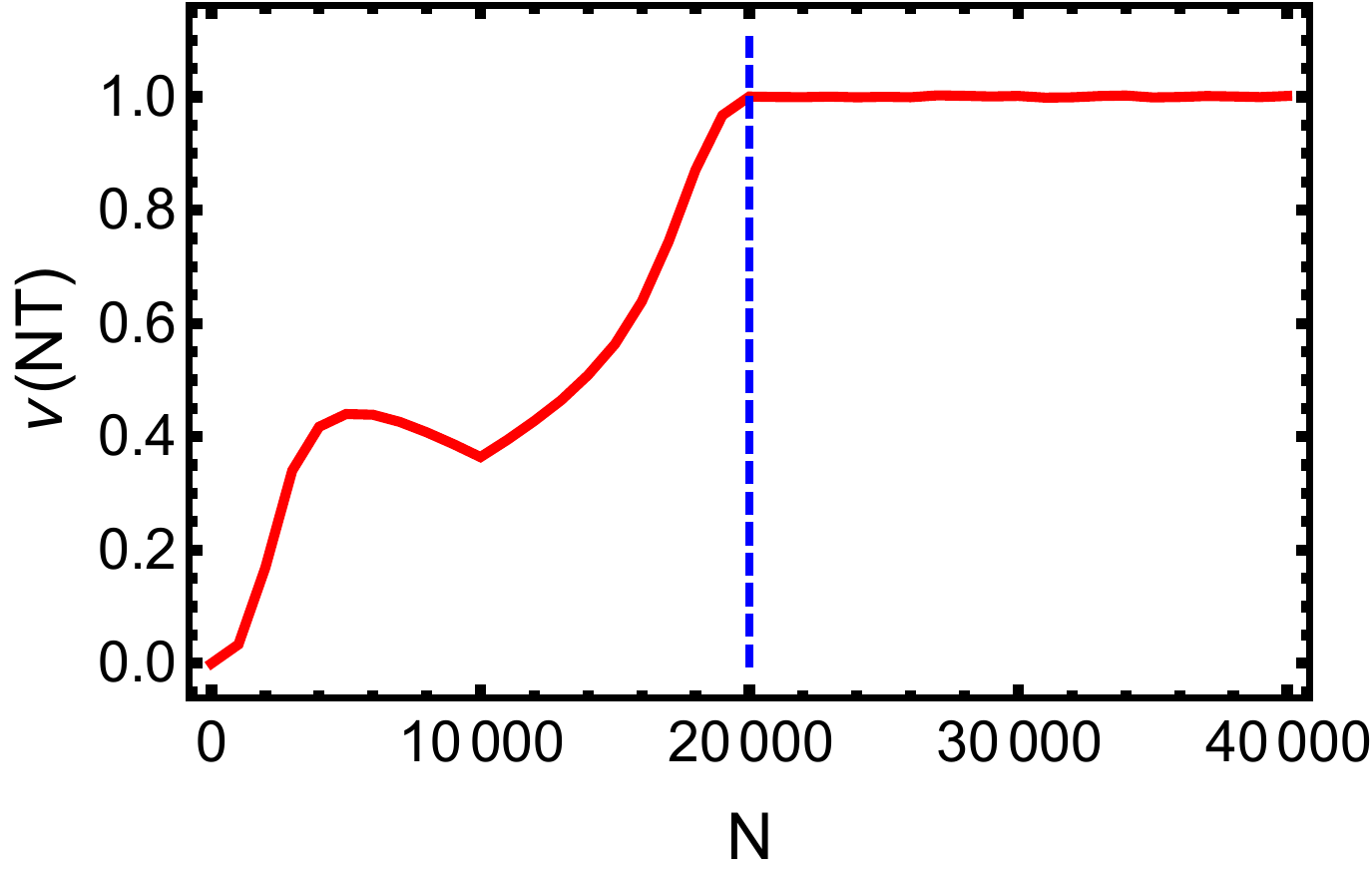}
		
		\caption{ Temporal variation of the winding number \ct{comment} as a function of stroboscopic time period for a periodic boundary condition. The winding number assumes a constant value (unity) after the $\mathcal{P}$ symmetry have been completely restored in the Floquet Hamiltonian at $t=2\tau$, with $\tau=10^4T$. The corresponding change in the bulk polarisation current is shown in the Appendix \ref{sec:current}.
		 \label{1a}  }
	\end{center}

\end{figure}

  The winding number calculated from the non-equilibrium state of the driven chain is also observed to become topologically non-trivial (see Fig.~\ref{1a}). In response to the explicit breaking of the $\mathcal{P}$ symmetry and the consequent variation of the winding number, we observe a stroboscopic generation of polarisation current in the bulk (see Appendix \ref{sec:current}) which vanishes just after the BDI symmetries are restored in the Floquet Hamiltonian at $t=2\tau$. Since the  $\mathcal{S}$ symmetry has been restored finally, the winding number also remains invariant further \ct{comment} and  the final Floquet Hamiltonian indeed becomes topologically inequivalent to the trivial Hamiltonian. The mass term necessarily avoids the crossing of a gapless QCP and therefore under an adiabatic protocol, the instantaneous state of the system $\ket{\psi_k(2\tau)}$ is seen to dynamically follow the ground state $\ket{\psi_0^F(k)}$ of the high frequency Floquet Hamiltonian.\\
  
\begin{figure}[h]
	\begin{center}
		\includegraphics[width=0.48\textwidth,height=0.65\columnwidth]{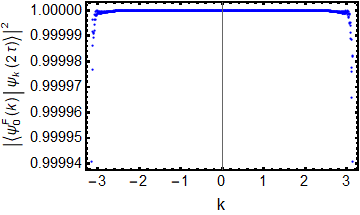}
		
		\caption{The overlap of the evolved state $\ket{\psi_k(2\tau)}$ at $t=2\tau$ with the Floquet eigenstate at the same time $\ket{\psi_0^F(k)}$ for all $k\in \left[-\pi,\pi\right]$.  For a sufficiently large $\tau$  ($\tau=10^4 T$), the instantaneous state of the system has a significant overlap with the Floquet topological ground state in the high frequency limit. The initial parameters are $v=1.55$, $w=1.50$ and $L=1000$; $\omega=100.0$ and amplitudes $V_0=0.1$, $V_1=7.0$ as in Eq.~(11) and Eq.~(13) of the manuscript.
			\label{fidelity}
		}
	\end{center}
\end{figure}

 {The adiabatic ramping functions $A(t)$ in  Eq.~(12) have been chosen so that its rate of change is much slower than the time period of the periodic perturbation (i.e., $\tau \gg T$). This ensures that the dynamical state of the system at stroboscopic intervals of time have a significant fidelity to the topological eigenstate of the Floquet Hamiltonian. In Fig.~\ref{fidelity}, we see that at $t=2\tau$, the stroboscopic state $\ket{\psi_k(2\tau)}$ has a significant overlap with the eigenstate $\ket{\psi_0^F(k)}$  of the Floquet}
  {Hamiltonian for all $k\in\left[-\pi,\pi\right]$.}

\section{Stroboscopic bulk boundary correspondence: OBC}
\label{sec: OBC}
 
 In this section we study the time periodic driving protocol given in Eq.~\eqref{protocol} in a SSH chain with open boundary conditions. The static SSH Hamiltonian $H(v_0,w)$ as in Eq.~\eqref{eq:H}, with $\mathcal{N}$ unit cells under open boundary conditions (i.e., $c^{\dagger}_{2\mathcal{N}+1}=0$), is a $2\mathcal{N}\times2\mathcal{N}$ hermitian matrix in the single-particle basis $\mathcal{B}\equiv\{\ket{i}\}_{i=1}^{2\mathcal{N}}$. Here $\ket{i}$ denotes the $i^{th}$ site in the chain.
 We start with an eigenstate $\ket{\psi(0)}$ of the initial topologically trivial Hamiltonian $H(v_0,w)$ having $\mathcal{N}=50$ unit cells and adiabatically switch on an unitary time dependent periodic perturbation,
\begin{equation}
H(t)=H(v_0,w)+v(t)\left(\sum_{i=1}^{2\mathcal{N}}c^{\dagger}_{n,1}c_{n,2}+h.c\right),
\end{equation}
where,
\begin{equation}\label{drive}
\begin{split}
v(t)= - \frac{t}{\tau}\left[V_0+V_1 e^{-i\omega t} \right] \text{~~for~~} 0\le t \le \tau , \\
v(t)=  -V_0 -\left(1-\frac{t-\tau}{\tau}\right)V_1 e^{-i\omega t}  \text{~~for~~} \tau\le t \le 2\tau,\\
v(t)= -V_0 ~~~~\text{~~for~} t\ge 2\tau, \\
\end{split}
\end{equation}
and $\omega=2\pi/T$ is the frequency of the drive while $T$ being it's time period.
 The matrix elements of the dynamical unitary propagator $U(t)$ in the time-independent basis $\mathcal{B}$ under the protocol Eq.~\eqref{drive}, satisfies a linear system of 
  \begin{figure}
 	\includegraphics[width=0.47\textwidth,height=0.65\columnwidth]{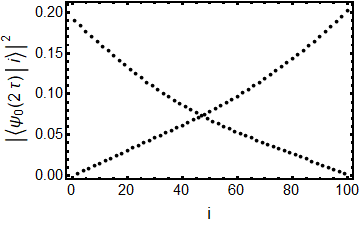}
 	\caption{Under open boundary conditions, at $t=2 \tau$, the stroboscopic time evolved state shows localisation at the ends of the chain after the BDI symmetries have been restored in the topological Floquet Hamiltonian. The energy of the localised state $E=0.003$ for a finite chain with 100 lattice sites. The initial state (same as in (a)) is subjected to a unitary drive as in Eq.~\ref{protocol} with $V_0=0.1$, $V_1=2.0$, $\tau=500T$ and $\omega=100.0$, where $i$ denotes the lattice site.  }
 	\label{2b}
 \end{figure}
  $2\mathcal{N}\times 2\mathcal{N}$ differential equations,
 \begin{equation}\label{schrodinger}
 i\frac{dU(t)}{dt}=H(t)U(t).
 \end{equation}
 To find the stroboscopic propagator $U(nT,(n-1)T)$, which dynamically evolves the state of the system from the time $t=(n-1)T$ to $t=nT$, we iteratively solve Eq.~\eqref{schrodinger} with appropriate boundary conditions in each interval $\left[(n-1)T,nT\right]$.
 By a repetitative application of the unitary propagator $U(nT,(n-1)T)$ for each stroboscopic interval, we obtain the time-evolved state of the system at $t=NT=2\tau$,
 \begin{equation}
\ket{\psi(2\tau)}=\prod_{n=1}^{N=\frac{2\tau}{T}} U(nT,(n-1)T)\ket{\psi(0)}
 \end{equation}
 
 To maintain adiabaticity of the deformation of the Floquet Hamiltonian   {which is essential for the adiabatic preparation}, we choose $\tau \gg T$ ($\tau=500T$) and a sufficiently high freqency $\omega=100$ for the periodic perturbation. The initial parameters of the topologically trivial SSH chain $H(v_0,w)$ were chosen to be $v_0=1.55$ and $w=1.50$. With the choice of the amplitudes $V_0=0.1$ and $V_1=2.0$ as specified in the   (see Fig.~\ref{2b}), we study the support of the stroboscopic state $\ket{\psi(2\tau)}$ of the system at the $i^{th}$ site of the chain vide the quantity $|\langle i\ket{\psi(2\tau)}|^2$. We observe localisation of the state $\ket{\psi(2\tau)}$ to the two ends of the finite chain corresponding to the topologically non-trivial winding number of the bulk Floquet Hamiltonian depicted in Fig.~\ref{1a}.
 We note that although the generated stroboscopic end modes de-localise considerably into the bulk of the
   \begin{figure*}
 	\begin{center}
 		\subfigure[]{\label{3a}}{\includegraphics[width=0.48\textwidth,height=0.8\columnwidth]{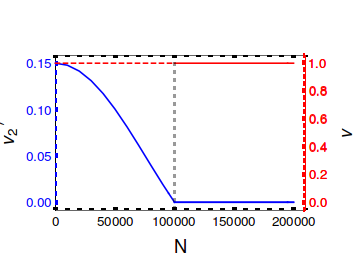}}
 		\subfigure[]{\label{3b}}{\includegraphics[width=0.44\textwidth,height=0.63\columnwidth]{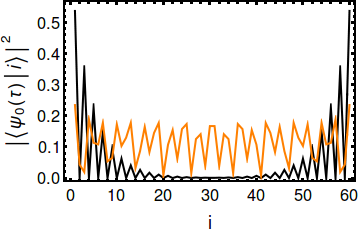}}
 		
 		\caption{ (a) Under the time-periodic drive (as discussed in Sec.~\ref{sec: SecondHopping} ), the renormalised second neighbour hopping strength $v_2^{\prime}$ as in Eq.~\eqref{eq_v2prime} (shown in blue) of the extended SSH chain eventually vanishes as the drive is gradually switched on. As a result, the BDI symmetries are restored in the Floquet Hamiltonian. The dynamical winding number (shown in red) however remains quantised to unity stroboscopically throughout. For the initial  Hamiltonian $H^E(k)$ in Eq.~\eqref{extended_momentum},  the parameters are  $v=0.15$, $w=0.22$, $v_2=0.15$ $v_3=0$ and $w_3=0.015$;  further we have chosen  $\omega=10^8$, $\tau=10^5T$ and $A_0=1.2$, and a  system size of $L=1000$.
 			(b) The corresponding stroboscopic state $\ket{\psi_0(\tau)}$ of the system eventually hosts zero energy states (under OBC with a system size $L=60$) localised at the ends of the chain (shown in black)  after the restoration of the BDI symmetries in the Floquet Hamiltonian for $t\geq\tau$. This establishes that starting from a delocalised eigenstate $\ket{\psi_0(0)}$ (shown in orange) of the extended SSH chain, the vanishing of the stroboscopically renormalised second hopping restores the stroboscopic bulk-boundary correspondence in the dynamical state of the system.				}
 	\end{center}
 \end{figure*}
  chain, in the thermodynamic limit the hybridization between the end modes will eventually vanish and they will get strictly localised at the edge.

  The stroboscopic state thus prepared at $t \geq 2\tau$, is observed to host topologically protected zero energy modes localized at the ends of a finite chain with open boundary conditions (see Fig.~\ref{2b}). As the parameters of the final Floquet Hamiltonian are close to the critical values, the localization length of the zero energy edge states are large. The end states therefore hybridise near the midpoint of the chain which consequently lifts the energy of the localized states from exactly zero;  this is also true  for an undriven SSH chain for the same set of parameters.  However,
  for a sufficiently long chain,  the edge states formed would be highly localised.   \\
  
  \section{Extended SSH chain and topology}
  \label{sec: extendedSSH}
   {To further exemplify the unitary protocol studied in Sec.~\ref{sec: protocol}, we now consider an extended  SSH chain with  additional second and third nearest neighbour hopping terms. In  particular, the presence of the second neighbour hopping  explicitly  {breaks the BDI symmetries of} the SSH Hamiltonian and consequently the edge modes are no longer protected against delocalisation into the bulk of the chain. The Hamiltonian describing the extended SSH chain \ct{gozalez18},
  \begin{equation}\label{extended}
  \begin{split}
  H^E=\sum_{i,j=1}^{2\mathcal{N}}t_{ij}c^{\dagger}_ic_j=H_{SSH}(v,w)\\
  +v_2\sum_{i=1}^{\mathcal{N}}\left(c_{i,A}c_{i+1,A}^{\dagger}+c_{i,B}c_{i+1,B}^{\dagger}\right)\\
  +\sum_{i=1}^{\mathcal{N}}v_3c_{i,A}c^{\dagger}_{i+1,B}+w_3c_{i,B}c^{\dagger}_{i+2,A}+\text{H.c.}
  \end{split}
  \end{equation}
  where H.c. denotes the hermitian conjugate and $H_{SSH}$ represents the SSH Hamiltonian given in Eq.~\eqref{eq:H}. Under PBC, the Hamiltonian in Eq.~\eqref{extended} can be recast in the decoupled form in the momentum basis,
  \begin{equation}
  H^E(k)=h_0(k)\mathbb{I}+\vec{h}(k).\vec{\sigma}
 \label{extended_momentum}
  \end{equation}
  where,}

  {\begin{equation}\label{ex_hk}
  \begin{split}
  h_0(k)=2v_2\cos(k),\\
  h_x(k)=v+(w+v_3)\cos{k}+w_3\cos{2k},\\
  h_y(k)=(w-v_3)\sin{k}+w_3\sin{2k},\\
  h_z(k)=0.
  \end{split}
  \end{equation}}
 	
 	\begin{figure*}
 	\begin{center}
 		\subfigure[]{\label{4a}}{\includegraphics[width=0.46\textwidth,height=0.63\columnwidth]{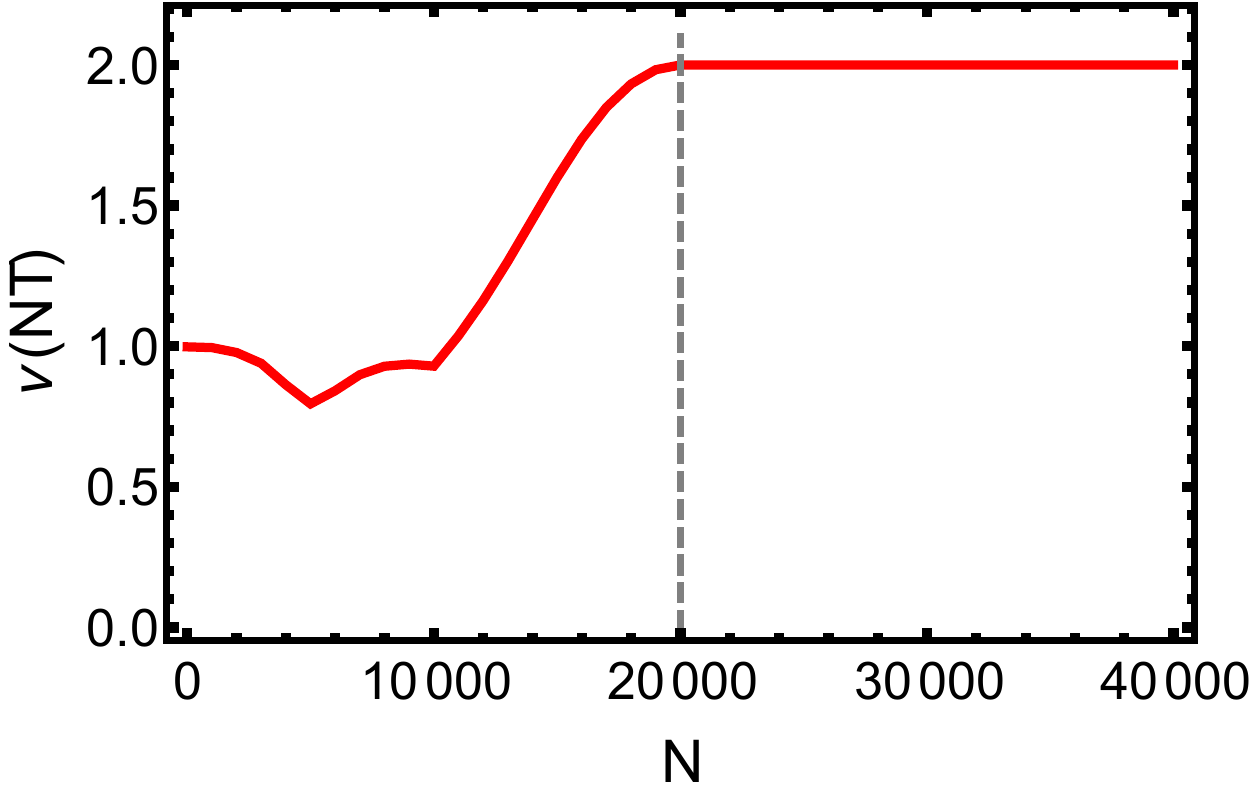}}
 		\subfigure[]{\label{4b}}{\includegraphics[width=0.46\textwidth,height=0.63\columnwidth]{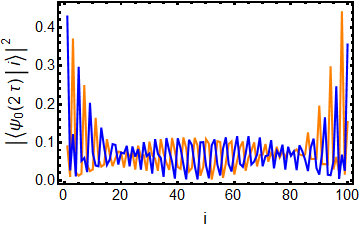}}
 		
 		\caption{ (a) Temporal variation of the stroboscopically observed dynamical winding number (under the unitary driving protocol discussed in Sec. \ref{winding2}), starting from an eigenstate of the BDI symmetric extended SSH Hamiltonian with the second neighbour hopping $v_2=0$   in Eq.~\eqref{extended_momentum}. We further choose  $v=1.0$, $w=2.0$, $v_3=0$ and $w_3=0.5$  so that  the  equilibrium winding number of the initial state is unity. In the presence of the time-dependent drive as in Eq.~\eqref{onetwo} with $V_0=-1.0$, $V_1=7.0$, $\tau=10^{4}T$ and $\omega=100$,  the dynamical winding number eventually assumes a stroboscopically quantised value of two. The winding number remains invariant thereafter for $t\geq2\tau$ i.e. when the $\mathcal{P}$ symmetry has been completely restored in the Floquet Hamiltonian. The system size $L=1000$.
 			(b) Under OBC, the out-of-equilibrium state of the driven extended SSH chain hosts two zero energy states localised at different sites of the chain (shown in blue and orange). Starting from an eigenstate of the Hamiltonian in Eq.~\eqref{extended} with $v=0.5$, $w=0.1$, $v_2=0$, $w_3=0.25$, $v_3=0$ and for a system size $L=100$, we follow the corresponding unitary protocol specified in Sec.~\ref{winding2}. The drive is chosen such that,$V_0=-0.5$, $V_1=2.0$, $\tau=10^3T$ and $\omega=100$. This establishes a stroboscopic bulk-boundary correspondence once the $\mathcal{P}$ symmetry has been completely restored in the Floquet Hamiltonian even for higher dynamical winding number state.				}
 	\end{center}
 \end{figure*}
 
   {The presence of the term proportional to $\mathbb{I}$ in the Hamiltonian in Eq.~\eqref{extended_momentum} breaks both $\mathcal{P}$ and $\mathcal{S}$ symmetries explicitly in the Hamiltonian while preserving the time reversal invariance if the hopping strengths are chosen to be real. According to the topological characterization of non-interacting systems \ct{altland97}, the model belongs to the AI class, which is trivial in 1D. However, the second neighbour hopping preserves the space-inversion symmetry which causes the winding number to remain well-defined and quantised as the number of complete windings of the vector $\vec{h}(k)$ for $k\in\left[-\pi,\pi\right]$; it can assume integer values $-1$ to $2$ depending the hopping strengths \ct{gozalez18}.}\\
 
  {In the absence of the third neighbour hopping ($v_3=w_3=0$), the second neighbour hopping strength distinguishes the band structure into two very different physical regimes (which however, are not differentiated by a topological phase transition): (i) if $v_2\le v/2$, the bulk of the chain is still in an insulating phase with a direct band-gap. However, even if the winding number stays non-zero and quantised, the energies of the localised end modes gradually move away from zero and they delocalise into the bulk of the chain as the second neighbour hoping strength increases above zero. (ii) if $v_2\ge v/2$, the bulk bands develop an overlapping indirect gap without crossing each other and the maxima of the valence band becomes higher in energy than the minima of the conduction band. The bulk behaviour of the system becomes metallic and consequently the end states delocalise completely into the bulk of the chain irrespective of the bulk winding number, thus breaking the bulk-boundary correspondence (see Fig.~\ref{3b}).}

  \subsection{Dynamical restoration of stroboscopic bulk-boundary correspondence} \label{sec: SecondHopping}
 { In this section, we start from an eigenstate of the Hamiltonian in Eq.~\eqref{extended} with non-zero strengths of the extended second neighbour and third neighbour hoppings 
 such that the initial Hamiltoinan is in the AI class and the edge states are completely delocalised in the bulk.  We have further chosen $v_2 >v/2$ so that initially the system is in the
 metallic state.}
 
  {A perfectly periodic perturbation is switched on adiabatically such that all the hopping parameters of the time-dependent Hamiltonian $H_k^E(t)$ are temporally modulated,}

  {\begin{equation}
 \label{eq:bessel_1}
 \begin{split}
 v(t)\rightarrow ve^{iV(t)},~~ w(t)\rightarrow we^{iV(t)},\\
 v_2(t)\rightarrow v_2e^{2iV(t)},~~v_3(t)\rightarrow v_3e^{3iV(t)},\\
 w_3(t)\rightarrow we^{3iV(t)},
 \end{split}
 \end{equation} }

 { where,
 \begin{equation}\label{eq:ramp_ext}
 \begin{split}
 V(t)=\frac{A_0t}{\tau}\sin{\omega t},~\text{for}~0\leq t\leq \tau\\
 =A_0\sin{\omega t}~\text{for}~t\geq \tau.
 \end{split}
 \end{equation}}

  {The ramping time $\tau$ is chosen in such a way that $\tau\gg T$, where $T=2\pi/\omega$ is the time period of the periodic part of the drive (as discussed in Sec.~\ref{sec: protocol}). Also, under the high frequency approximation ($\mathcal{A}^2/\omega\ll 1$,where $\mathcal{A}$ is the {\it effective} amplitude of the drive at any time), the stroboscopic dynamics of the chain is effectively described by the Floquet Hamiltonian in the zero photon sector,
 \begin{equation}\label{eq:floquetavg}
H_F^{t,E}(k)\simeq\frac{1}{T}\int_0^TH_k^E(t^{\prime})dt^{\prime}=h_0(t)^{\prime}(k)\mathbb{I}+\vec{h}^{\prime}(t)(k).\vec{\sigma}
 \end{equation}
 such that, $h_i^{\prime}(t)$ take similar forms as in Eq.~\eqref{ex_hk} with renormalised hopping strengths
 \begin{equation}\label{bessel_2}
 t^{\prime}_{ij}(t)\rightarrow t_{ij}\mathcal{J}_0\left(\frac{A_0t}{\tau}d_{ij}\right),
 \end{equation}
 where we have assumed that $\tau\gg\ T$ so that the amplitude remains approximately constant in one period of the sinusoidal drive. $\mathcal{J}_0$ is Bessel function of the zeroth order and $d_{ij}$ is the distance between the two lattice points $i$ and $j$. }
 
 {We now choose the amplitude $A_0$  such that the stroboscopically renormalised second neighbour hopping 
\be
v^{\prime}_2 = v_2 \mathcal{J}_0\left(\frac{2A_0t}{\tau}\right) 
\label{eq_v2prime},
\ee
(with lattice spacing $a=1$) vanishes completely (so that the $h_0^{\prime}(k)$ in Eq.~\eqref{ex_hk} vanishes) when the time-dependent perturbation has been switched on completely, i.e., at $t=\tau$. This restores the BDI symmetries in the Floquet Hamiltonian. The adiabatic protocol further ensures that the stroboscopic state of the system $\ket{\psi_0(\tau)}$ inherits all the BDI  symmetries and thus becomes topological; this can be verified using a similar fidelity analysis for the present case as presented in Fig.~\ref{fidelity} for
the SSH model. Considering the same protocol, with an OBC, one finds stroboscopic zero-energy modes localised at the edge of the chain. In Figs.~\ref{3a}-\ref{3b}), we show  the stroboscopic variation 
of the winding number and the emergence of corresponding edge states at $t \geq\tau$.}

 {Thus,  the gradual decrease in the stroboscopically renormalised second hopping modifies the bulk Floquet bands and drives the system from an AI symmetric  metallic phase to a BDI symmetric topological insulator phase and consequently the bulk-boundary condition is stroboscopically restored in the out-of equilibrium state of the system.}
  
  \subsection{Unitary preparation of higher winding number states}
  \label{winding2}
 {As was discussed in Sec.~\ref{sec: SecondHopping}, the coherent destruction of the second neighbour hopping restores the BDI symmetries in the out of equilibrium stroboscopic state of the system. In this section, we set the second neighbour hopping strength $v_2$ to zero at the outset.
 {The initial state is chosen} to be an eigenstate of a BDI symmetric extended SSH Hamiltonian with hopping strengths $(v,w,v_3,w_3)$ as in Eq.~\eqref{ex_hk} characterised by  a winding number of unity. The corresponding
two-level Hamiltonian  $H^E_k(v,w,v_3,w_3)$ has no second nearest hopping (i.e., $h_0(k)=0$) but a non-zero third nearest hopping.\\}

 {Our aim here is to dynamically prepare a topological state with stroboscopic winding number two starting from the state with equilibrium winding number unity. Through a similar unitary driving protocol as discussed in Sec.~\ref{sec: protocol}, we adiabatically ramp a time-periodic perturbation to prepare the system in a topological Floquet state.
The unitary protocol,
\begin{equation}\label{onetwo}
H_k^E(t)=H^E_k\left(v,w,v_3,w_3-A_1(t)V_0\right)+A_2(t)V_k(t),
\end{equation}
where $H_k^E$ is the extended SSH Hamiltonian with time-dependent hopping amplitudes and,
\begin{equation}\label{protocol_ex}
\begin{split}
A_1(t)=A_2(t)= -t/\tau\text{~~for~~} 0\le t \le \tau, \\
A_1(t) =1 \text{~~and~~} A_2(t) = -\left(1-\frac{t-\tau}{\tau}\right) \text{~~for~~} 	\tau\le t \le 2\tau,\\
A_1(t) =1 \text{~~and~~} A_2(t) = 0 \text{~~for~~} 	t\ge 2\tau.
\end{split}
\end{equation}
$V_k(t)$ is a perfectly periodic perturbation of the form specified in Eq.~\eqref{Vkt}.  As discussed in Sec. \ref{sec: protocol}, during the evolution from $0\le t \le \tau$, the chiral mass term is adiabatically switched
on while from $\tau\le t \le 2\tau$, it is adiabatically switched off  (see Eq.~\eqref{eq:floquet}).  The periodic perturbation $V_k(t)$ thus induces a gap in the Floquet Hamiltonian during the entire adiabatic
ramping process, thereby protecting the dynamical system against the generation of excitations. Under the complete unitary protocol in Eq.~\eqref{protocol_ex}, the stroboscopic state of the system makes a transition from an initial  state having a winding number of unity to a higher winding phase having winding number two for stroboscopic times $t\geq 2\tau$. Owing to the adiabaticity of the switching on of the periodic perturbation, it can be verified that the out-of equilibrium stroboscopic state of the system maintains a high fidelity with the Floquet topological state.}

 {Consequently, the stroboscopically observed extended chain under OBC, exhibits the dynamical emergence of two pairs of localised edge states in contrast to a single pair of edge states in the initial equilibrium system. Thus, we establish a stroboscopic bulk boundary correspondence even in a out-of-equilibrium state with higher winding number (see Fig.~\ref{4a}-\ref{4b}). }

\section{Discussions and experimental connections}
\label{conclusion}

 Exploiting the possibility of the temporal variation of the winding number
   in a 1D topological insulator 
   under a unitary drive,
we have established the possibility of engineering  the state of a 1D topological insulator in its topologically non-trivial phase.
 We start from a trivial ground state of a BDI symmetric SSH chain and let it evolve under the application of a time periodic perturbation whose amplitude is adiabatically modified such that
it may be assumed to be nearly constant within a complete period of the drive. The complete ``switching on" of the periodic potential drives the parameters of the Floquet Hamiltonian from a topologically trivial to a topologically non-trivial value together with the generation of a chiral mass in $H_F(k)$. Subsequently, the periodic drive generating the staggerred mass term in the Floquet Hamiltonian is switched off adiabatically to  restore the BDI symmetries in the Floquet Hamiltonian. As no gapless QCP is crossed, the instantaneous state of the system (nearly) follows the ground state of the Floquet Hamiltonian.

We have also established the dynamical emergence of topologically protected localized zero energy modes in the stroboscopic state of a long SSH chain under open boundary conditions. Our study, therefore, unravels the possibility of the unitary preparation of a 1D Floquet topological phase through an adiabatic deformation of the Floquet Hamiltonian. \\

 {Thereafter, we have unitarily prepared a topologically non-trivial stroboscopic state starting from a trivial equilibrium state in an extended SSH chain with higher neighbour hoppings. The extended SSH Hamiltonian having finite second and third neighbour tunnelling strengths belong to the AI symmetry class of non-interacting topological systems and is topologically trivial in 1D. Furthermore,  for a sufficiently high second neighbour hopping strength the bulk of the chain is in a metallic phase with an indirect band-gap. Starting from such an indirect band-gap phase we adiabatically switch on a time-periodic perturbation to stroboscopically restore the BDI symmetries in the system and prepare it into a stroboscopic topological insulator state. Consequently, the out of equilibrium stroboscopic state of the system exhibits a dynamical emergence of localised edge states under OBC. This establishes the existence of an out-of-equilibrium bulk boundary correspondence in periodically driven 1D systems.\\}
 {Lastly, starting from a BDI symmetric extended SSH chain with vanishing second neighbour hopping but a finite third neighbour tunnelling (having a winding number of unity), we unitarily prepare the system in a stroboscopic state having a winding number two. The stroboscopic state of the corresponding chain under OBC exhibits two sets of topologically protected localised edge states, thereby establishing the existence of a non-equilibrium bulk-boundary correspondence even in higher winding phases.}

The Floquet Hamiltonian approach is fundamentally important experimentally \ct{atala13} as the exact simulation of tight-binding Hamiltonians in ultracold atomic setups are essentially achieved through the application of time-periodic fields.
Furthermore, the preparation protocols of a topological state under adiabatic quenching of a parameter in the Hamiltonian, requires the dynamical process to be infinitely slow compared to the relaxation time of the system which is inversely proportional to its bandwidth. In the Floquet adiabatic deformation approach, however, the rate of deformation of the Floquet Hamiltonian only needs to be sufficiently slower than the frequency of the periodic part which may be chosen to be large enough so that the Floquet band-gap is large. This allows one to reach the ``stroboscopic" topological state within an experimentally realizable time scale although the Floquet approach sacrifices the topological non-triviallity in the micromotion.
The survival of the stroboscopic bulk-boundary correspondence under a time-periodic drive, provides with an experimentally feasible way to dynamically engineer topological properties in non-equilibrium systems.  These considerations emphasises the importance of our time-periodic protocol.  {We note that if the complete driving protocol is tuned to be adiabatic, the stroboscopic change in the winding number will be quantised as in topological charge pumping in a periodically driven Rice-Mele model \ct{asboth16}.}

We conclude with the note that, the bulk topological invariant of 1D topological systems has  been probed experimentally using ultracold atomic setups in optical lattices \ct{atala13}. The recent experimental study in optical lattice of trapped $^{87}Rb$ atoms \ct{meier16} has established the experimental preparation of topologically non-trivial states.  {The extended SSH model has also been experimentally realised recently, with ultracold atoms in momentum space to establish its topological properties and bulk-boundary correspondence \ct{xie19}.} Robust edge states were also observed and extensively studied in synthetically prepared {\it Floquet} topological systems in ultracold atomic lattices \ct{reichl14,velasco17}. This is one of the fundamental motivations behind the choice of a periodic drive in the proposed dynamical protocol to engineer the topological state. The bulk polarisation current  {(discussed in Appendix \ref{sec:current})} and the stroboscopically localised topological edge states, both being observable quantities are expected to be measurable in similar experimental setups in optical lattices.

\appendix

\section{The Bulk polarisation current in the SSH model:}
\label{sec:current}

 It has also been established \ct{bandyopadhyay19} that under an arbitrary time-dependent driving, the bulk polarization current density $j(t)$ of the SSH chain is directly proportional to the rate of change of 
 the topological winding number $(\nu)$:
\begin{equation}\label{w_j}
j(t)=\frac{1}{2\pi}\int_{BZ}\bra{\psi_k(t)}\partial_{k}H_k(t)\ket{\psi_k(t)}=\frac{d\nu}{dt},
\end{equation}
{where $H_k(t)$ is the driven instantaneous Hamiltonian. In the presence of a time-dependent driving which breaks the particle hole symmetry in the dynamics, a finite polarization current is generated in the bulk of the chain. The unitary protocol conceived in Sec.~\ref{sec: protocol} of the main text explicitly breaks the $\mathcal{P}$ symmetry  in the $H_F^t(k)$ for all stroboscopic times earlier to $t=2\tau$. Consequently, one sees a stroboscopic generation of a bulk polarization current which eventually vanishes at $t\geq2\tau$ when all the BDI symmetries have been restored in the Floquet Hamiltonian (see Fig.~\ref{fig:A1}).
\begin{figure}[h]
	\includegraphics[width=0.48\textwidth,height=0.655\columnwidth]{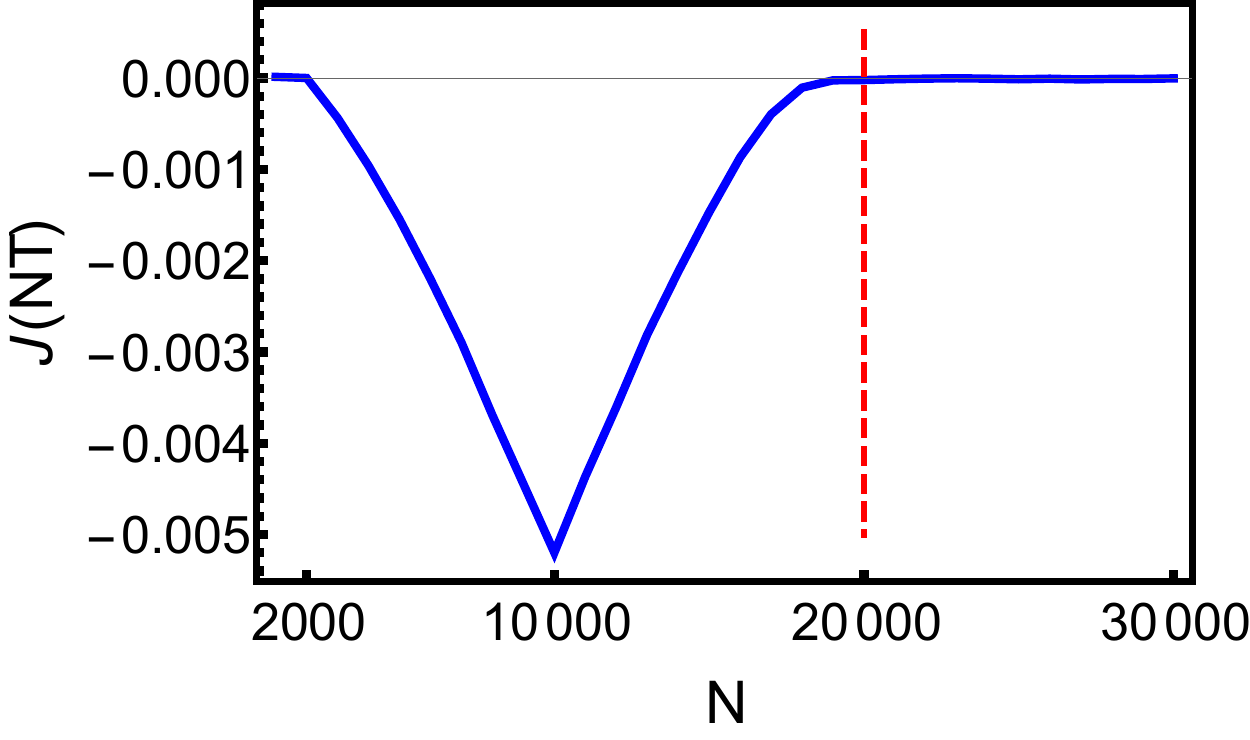}
	
	\caption{ 
		As a result of the variation of the winding number,  the stroboscopic bulk polarisation current vanishes for $t >2\tau$. The initial state is the ground state  of  a  BDI symmetric SSH model with parameters $v=1.55$, $w=1.50$ and $L=1000.$ We have
		chosen $\omega=100.0$ and amplitudes $V_0=0.1$, $V_1=7.0$ as in Eq.~\eqref{protocol}. The variation of the polarisation current perfectly conforms to the variation of the dynamical winding number 
		presented in Fig.~\ref{1a} and vanishes when it reaches the value unity }
	\label{fig:A1}
\end{figure}
	
	In a periodically driven Rice-Mele chain, the stroboscopic current resembles a stroboscopic quantized transport of charge through the bulk (corresponding to a quantized cyclic change in the dynamical winding number) provided that the complete dynamics is adiabatic. However, for a generic symmetry breaking drive, the dynamical winding number varies continuously with time.

\begin{acknowledgements}
 We acknowledge Utso Bhattacharya, Sourav Bhattacharjee, Ritajit Kundu and Somnath Maity  for  discussions. We especially acknowledge Diptiman Sen
for critical comments. SB acknowledges CSIR, India for financial support.
\end{acknowledgements}


\end{document}